\documentclass[aps,prd,amsmath,showpacs,12pt]{revtex4}
\usepackage{epsfig}
\usepackage{graphics}
\usepackage{latexsym}
\usepackage{amsmath}
\usepackage{amssymb}
\usepackage{rotating}
\usepackage{subfigure}
\usepackage{bm}
\usepackage{color}

\usepackage[colorlinks = true,
linkcolor = magenta,
urlcolor  = blue,
citecolor = red,
anchorcolor = blue]{hyperref}
\newcommand{\beq}{ \begin{equation}}
\newcommand{\eeq}{\end{equation}}

\begin{document}
\title{
Properties of structure functions from helicity components\\
of light quarks and antiquarks in the statistical model}
\author{Claude Bourrely}
\email{claude.bourrely@cpt.univ-mrs.fr}

\affiliation{Aix Marseille Univ, Universit\'e de Toulon, CNRS, CPT, 
Marseille, France}

\begin{abstract}
In the quantum statistical parton distributions approach proposed more than one
decade ago to describe the parton structure, new properties are now understood,
in particular, the relation between quarks and antiquarks which leads
 to very specific properties.
The simultaneous treatment of unpolarized and polarized  Parton Distribution 
Functions (PDFs) allows a determination of thermodynamical potentials 
(the master parameters of the model) which drive their behavior and as a
consequence those of the structure functions.
The existence of a possible relation  between the gluon and a $q~\bar q$ pair
 leads to define a toy model for the unpolarized and polarized gluon.
In view of  forthcoming experimental results in the large $x$ region
specific predictions made by the model are presented.
\end{abstract}


 
\maketitle

\section{ Introduction}
\setcounter{page}{1}
Our main objective is to build  quarks structure where constitutive elements
can be  understood through their parameters which are easily associated
with the quark properties.
The first point to clarify is the choice of a statistical model. Taking 
the example of a proton at rest 
which contains three quarks a statistical treatment
seems not justified due to the 
low number of elements. 
However, when a proton is accelerated in a collider the energy
increase has not only an effect on  the masses but also to create
a large number of $q~\bar q$ pairs or a quark gluon plasma
where in a p-p collison materializes mainly
in primary unstable particles observed in a detector  by
 a large number of tracks.
The production of numerous pairs and gluons provide 
a justification for a statistical treatment of the partons interaction process.
Moreover, the fact that a quark is described in our model 
by a Fermi function means that it is
already dressed to live in a surrounding nuclear medium made of quarks.

The statistical approach is characterized by thermodynamical potentials
whose values are the master parameters, they drive not only the shape of the
PDFs but are found to control some specific properties of the structure functions.
In order to introduce the maximum  constraints one  decides to work
from the beginning with helicity components which are the building blocks
of both the polarized and unpolarized PDFs a unique situation in the domain.
It is clear that the number of polarized data is much smaller
than the unpolarized ones and also that they are limited to a medium energy region,
however the RHIC experiments enlarge somehow this domain but a large gap
remains to reach the LHC energy.

The objective of the paper is to discuss the consequence of the statistical
approach on the quarks structure because from a collection of results
obtained through the years one gets  now a better global view.

In section 2, a review of the basic elements of the quarks distribution is presented
together with the notations. In section 3, a proof is given to show how the 
antiquarks PDFs can be deduced from the quarks by using different 
constraints in the fitting process.
Section 4 is devoted to analyze the different helicity components and
how their effect is put in evidence in structure functions.
In section 5, a toy model is introduced to define new unpolarized
and polarized gluon PDFs, which is inspired by the relation between gluons
and $q \bar q$ pairs. The conclusions are presented in section 6.

\section{ Basic elements of the quark distributions}
The PDFs are the essential elements to evaluate
scattering processes in QCD. In the absence of a theory they are usually
parametrized with polynomials Refs \cite{ct14,mmht14}, to go beyond this 
approximation and in an attempt 
to define a more physical structure for the quarks a statistical approach 
was proposed many years ago to build up the PDFs \cite{bbs1}.

Let us now describe the main features of the statistical approach.
The fermion distributions are given by the sum of two terms,
a quasi Fermi-Dirac function of helicity $h = \pm$
and a helicity independent diffractive
contribution:
\begin{equation}
xq^h(x,Q^2_0)=
\frac{A_{q}X^h_{q}x^{b_q}}{\exp [(x-X^h_{q})/\bar{x}]+1}+
\frac{\tilde{A}_{q}x^{\tilde{b}_{q}}}{\exp(x/\bar{x})+1}~,
\label{eq1}
\end{equation}
for the quarks and for the antiquarks the ansatz:
\begin{equation}
x\bar{q}^h(x,Q^2_0)=
\frac{{\bar A_{q}}(X^{-h}_{0q})^{-1}x^{\bar{b}_ q}}{\exp
[(x+X^{-h}_{0q})/\bar{x}]+1}+
\frac{\tilde{A}_{q}x^{\tilde{b}_{q}}}{\exp(x/\bar{x})+1}~,
\label{eq2}
\end{equation}
at the input energy scale $Q_0^2=1 \mbox{GeV}^2$. 

With the above definitions the diffractive
term is the same for flavor $u, d$, but has a specific expression for other flavors. 
It is absent in the quark helicity distribution $\Delta q = q^+ - q^-$, 
 the quark valence contribution $q - \bar q$ and the difference $u - d$.\\
In the numerator of the non-diffractive parts of Eq.~(\ref{eq1}) 
the multiplicative factor $X^{h}_{q}$ allows to separate $u$ and $d$ quarks
since one  assumes $A_u = A_d$, the term $x^{b_q}$ imply a modification
of the quantum statistical form, this term is introduced in order to
control the small $x$ behaviour.
The parameter $\bar{x} = 0.09$ plays the role of a {\it universal temperature}
and $X^{\pm}_{q}$ are the two {\it thermodynamical potentials} of a quark
$q$, with helicity $h=\pm$. They represent the fundamental parameters of
the model because they  
drive the PDFs  behaviour\footnote{The PDF QCD evolution
was done at NLO in the $\overline {\bar {\mbox{MS}}}$ scheme  using the HOPPET 
program \cite{hoppet}.}. For convenience  the values of the potentials obtained
in BS15 \cite{Bourrely:2015kla} are recalled:

\begin{eqnarray}
\nonumber
X_{u}^+= 0.475 \pm 0.001,~X_{u}^-= 0.307 \pm 0.001,\\ 
X_{d}^+= 0.245 \pm
0.001,~ X_{d}^-= 0.309 \pm 0.001, \nonumber \\
X_{s}^+= 0.011 \pm 0.001,~X_{s}^-= 0.015 \pm 0.001.
\label{eq10}
\end{eqnarray}

\section{ Generation of the antiquarks distribution}
To adopt a coherent scheme it is natural to suppose that antiquarks
must also contain a Fermi part analogous to the quarks and also in addition a 
diffractive part being the same as in the quarks, 
all these constraints lead to a general expression like: 
\begin{equation}
x\bar{q}^h(x,Q^2_0)=
\frac{\bar A_q^{'h} x^{\bar{b}_q}}{\exp[(x-Y^{h}_{q})/\bar{x}]+1}+
\frac{\tilde{A}_{q}x^{\tilde{b}_{q}}}{\exp(x/\bar{x})+1}~.
\label{eq2a}
\end{equation}
This distribution depend on the new parameters $\bar A_q^{'h},  Y^{h}_{q}$
compared to Eq. (\ref{eq2}).
In order to determine these parameters in a fitting process the
constraint of the valence sum rule is added
\begin{equation}
\int (q(x) - \bar {q}(x))dx = N_q, \quad \mbox{where}\quad  N_q = 2, 1 
\quad\mbox{for~~ u, d}\,,
\label{valsr}
\end{equation}
(this sum rule is independent of the diffractive part) and
a second constraint which comes from the momentum sum rule
\begin{equation}
\int \sum_i  [xq_i(x) + x\bar {q}_i(x))+ xG(x)]dx = 1 \,,
\label{impuls}
\end{equation}
where $G(x)$ is the unpolarized gluon distribution.
Making a fit at NLO  of unpolarized and polarized experimental data 
analogous to the one discussed in BS15 Ref. \cite{Bourrely:2015kla} one finds
for the potentials a solution:
\begin{eqnarray}
&Y_u^-  = -0.475,\quad\quad Y_u^+ = -0.307 \nonumber \\
&Y_d^- = -0.244, \quad\quad Y_d^+ = -0.309\,,
\label{potval}
\end{eqnarray}
where a comparison with the solution obtained in BS15 (\ref{eq10}) leads to
\begin{equation}
Y_u^-  = -X_u^+,\quad Y_u^+ = -X_u^-,\quad Y_d^- = -X_d^+,\quad Y_d^+ = -X_d^- \,,
\label{potrelat}
\end{equation}   
the change of sign in the $\bar q$ potentials and in the helicity
find its origin from the unpolarized gluon whose potential is null
$X_q^{\pm} + Y_q^{\mp} = 0$, this point will be examined later.\\
The other parameters are given by:
\begin{eqnarray}
&&A = 1.943, \quad\quad b_u = b_d =0.471 \quad\quad \bar b_u = \bar b_d =1.304,\nonumber  \\
&&\bar A_u^{'+} = 29.039,\quad\quad \bar A_u^{'-} = 18.768 , \nonumber  \\
&& \bar A_d^{'+} = 28.851,\quad\quad\bar A_d^{'-} = 36.536, 
\label{param}
\end{eqnarray}
By introducing the definition $\bar A^{'h}_q = \bar A_q / X^{-h}_q $,
the antiquarks distributions (\ref{eq2a}) become identical to Eq (\ref{eq2}), where the
four normalizations $\bar A^{'h}_q$ are reduced to one constant $\bar A_q = 8.915$.
This result confirms the ansatz taken at the origin for the antiquarks, 
which was expected to be  a solution of Eq. (\ref{eq2a}).
To summarized an interesting relation between light quarks
and antiquarks in the statistical approach was established with the objective 
to reduce the number of arbitrary distributions (see sec. \ref{toymodel}).

\section{Properties of  the unpolarized and polarized quark distributions}
\label{propert}
From the results obtained in Eq. (\ref{eq10}) it is found for the light quarks
the following hierarchy between the different potential components
\begin{equation}
X_u^+ >  X_u^- \simeq X _d^- >  X_d^+ .
\label{potherar}
\end{equation}
In Eq. (\ref{potval}) the two potentials $X_u^-,X_d^- $
have very close numerical values, which is a consequence of the near equality 
between $xu^-(x, Q^2)$ and $xd^-(x, Q^2)$.

It is easy to show that quarks helicity PDFs increase with the potentials value, 
while for antiquarks helicity PDFs increase when the potentials decrease.

\begin{figure}[htp]   
\begin{center}
\includegraphics[width=6.5cm]{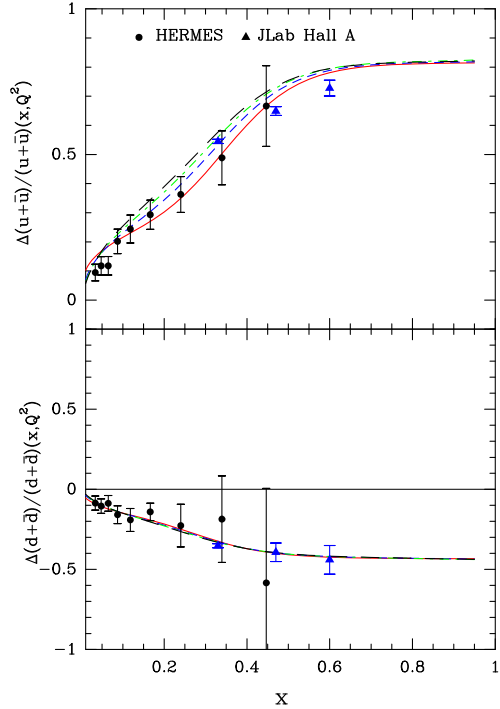}
\caption[*]{\baselineskip 1pt
BS15  \cite{Bourrely:2015kla}
predicted ratios $(\Delta u(x,Q^2) + \Delta \bar u(x,Q^2)) /  (u(x,Q^2 + 
\bar u(x,Q^2))$ and
$(\Delta d(x,Q^2) + \Delta \bar d(x,Q^2)) /  (d(x,Q^2 + \bar d(x,Q^2))$, versus $x$,
 at $Q^2 (\mbox{GeV}^2) =$  1 solid , 10 dashed , 100 dashed-dotted, 1000 long-dashed. 
Experiments: 
HERMES \cite{hermes1}-\cite{hermes3},  Jlab \cite{JLab1}-\cite{JLab2}.}
\label{disratiosl}
\end{center}
\end{figure}

As a consequence of the above hierarchy on potentials (\ref{potherar}) it follows 
a hierarchy on the quarks helicity distributions,
\begin{equation}
 xu_+(x) > xu_-(x) = xd_-(x) > xd_+(x)
\label{ineq}
\end{equation}
and a obvious hierarchy for the antiquarks, namely
\begin{equation}
x\bar d_- (x) >  x\bar d_+ (x) = x\bar u_+ (x) >  x\bar u_- (x) ,
\label{ineqbar}
\end{equation}
It is important to note that these inequalities Eqs. (\ref{ineq})-(\ref{ineqbar})
are preserved by the NLO QCD evolution. 
An other remark is the fact that the initial analytic form  
Eqs.~(\ref{eq1},\ref{eq2}), is almost preserved by the $Q^2$ evolution with 
some small changes on the parameters numerical values.
One clearly concludes that $u(x,Q^2) > d(x,Q^2)$ implies a flavor-asymmetric 
light sea, i.e. $\bar d(x,Q^2) > \bar u(x,Q^2)$, a trivial consequence of the Pauli 
exclusion principle, which is built in.
Indeed this is based on the fact that the 
proton contains two $u$ quarks and only one $d$ quark.\\ 
Let us move on to mention more significant consequences concerning the helicity
 distributions which follow from Eqs. (\ref{potval})-(\ref{ineqbar}).
First for the $u$-quark 
\begin{equation}
 x\Delta u(x,Q^2) > 0\,, \quad\quad  x\Delta \bar u(x,Q^2) > 0.
\end{equation}
Similarly for the $d$-quark 
\begin{equation}
 x\Delta d(x,Q^2) < 0\,, \quad\quad  x\Delta \bar d(x,Q^2) < 0\,,
\end{equation}
 these predictions were made almost 15 years ago \cite{bbs1}.
It is interesting to notice that the polarized structure function $xg_1^p$
measured by experiment and driven by $x\Delta u$
has a maximum around $ x = 0.42$  in a medium $Q^2$ range, such $x$ value
is  close to the thermodynamical potential $X_u^+$. 
Concerning $xg_1^n$ which is negative
for small $x$ because it is dominated by $x\Delta d$, when $x$ increases 
$x\Delta u$ becomes dominant so $xg_1^n$ takes positive values, all these
properties are well understood and described by the statistical model
due to the properties of thermodynamical potentials.
Our predicted signs and magnitudes have been also confirmed  
\cite{Bourrely:2015kla} by 
the measured single-helicity asymmetry $A_L$ in the $W^{\pm}$ production at 
BNL-RHIC from STAR experiment \cite{Adamczyk:2014xyw}.\\
Another important earlier prediction concerns the Deep Inelastic Scattering (DIS) 
asymmetries, more precisely 
$(\Delta u(x,Q^2) + \Delta \bar u(x,Q^2)) /  (u(x,Q^2 + \bar u(x,Q^2))$ and
$(\Delta d(x,Q^2) + \Delta \bar d(x,Q^2)) /  (d(x,Q^2) + \bar d(x,Q^2))$, 
shown in Fig. \ref{disratiosl}. 
Note that the  data from HERMES \cite{hermes1}-\cite{hermes3} and 
Jlab \cite{JLab1}-\cite{JLab2}, 
so far, are in agreement with these predictions at low $x < 0.6$. 
In the high $x$ region our prediction differs 
from those which impose, for both quantities, the value one for $x=1$. This is 
another challenge, since only up to $x=0.6$, they have  been measured at JLab
\cite{JLab1}-\cite{JLab2}.. 

\begin{figure}[htp]   
\begin{center}
\includegraphics[width=8.0cm]{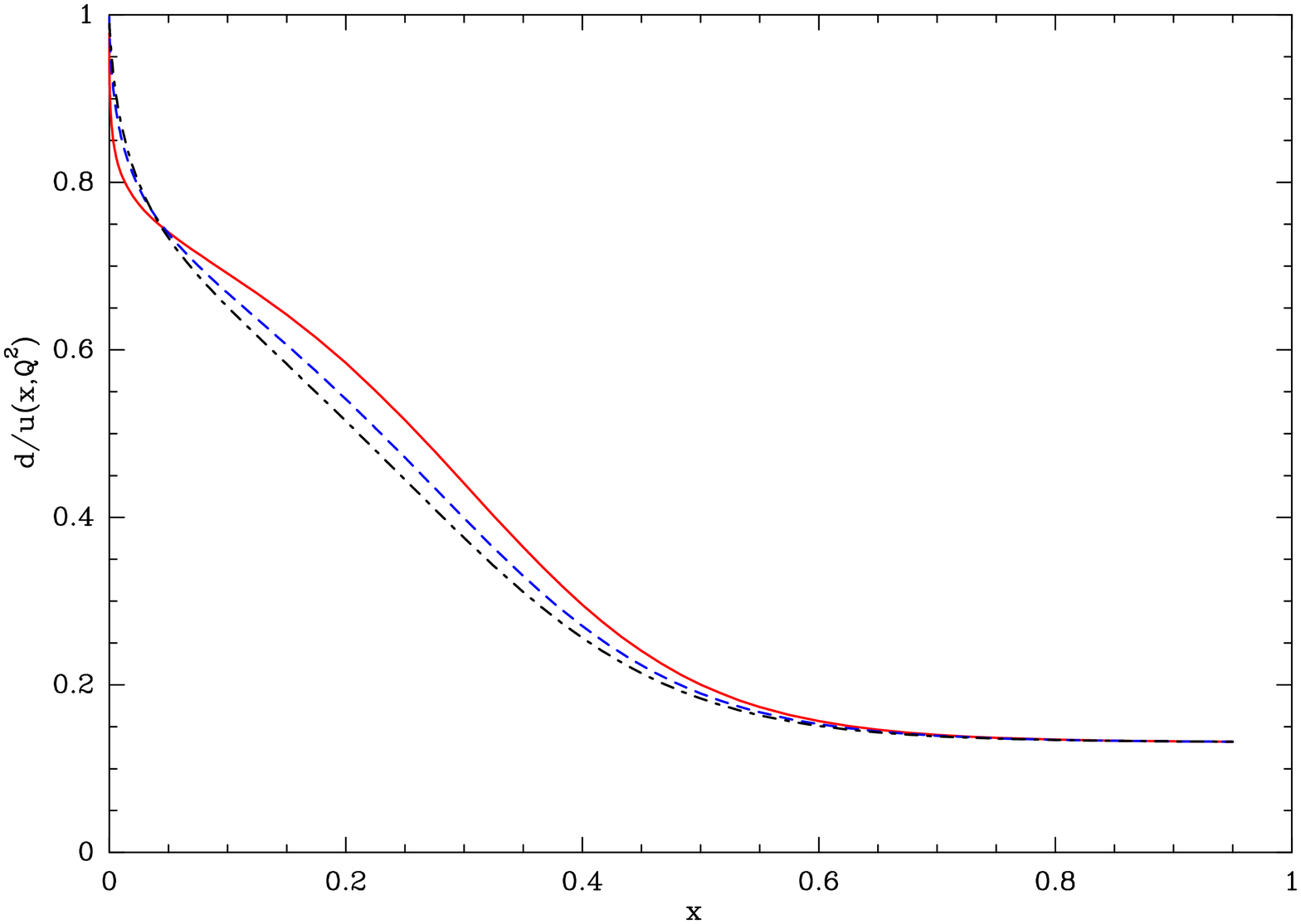}
\vspace*{+7.0ex}
\includegraphics[width=8.0cm]{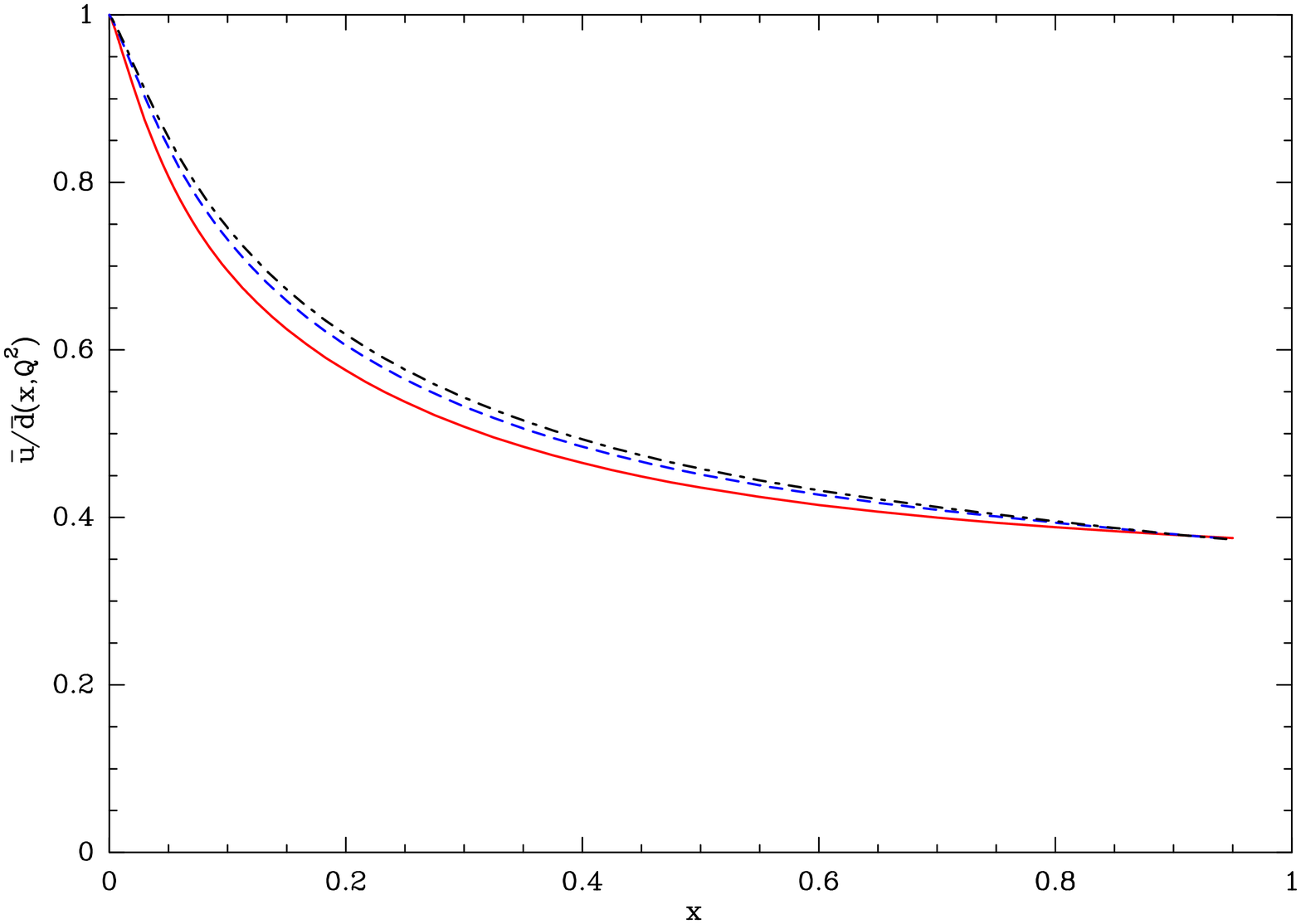}
\vspace*{-5.0ex}
\caption[*]{\baselineskip 1pt
The ratios $d(x,Q^2)/u(x,Q^2)$ (${\it left}$) and 
$\bar u(x,Q^2)/\bar d(x,Q^2)$ (${\it right}$) versus $x$ for
$Q^2 \mbox{GeV}^2 =$  1 solid , 10 dashed , 100 dashed-dotted,
from BS15  \cite{Bourrely:2015kla}.}
\label{ratios}
\end{center}
\end{figure}

There are two more important consequences which relate unpolarized and helicity 
distributions, namely for quarks
\begin{equation}
xu(x,Q^2) -  xd(x,Q^2) = x\Delta u(x,Q^2) -  x\Delta d(x,Q^2)   > 0 ,
\label{relatqdq}
\end{equation}
and similarly for antiquarks
\begin{equation}
x\bar d(x,Q^2) - x\bar u(x,Q^2) = x\Delta \bar u(x,Q^2) -  
x\Delta \bar d(x,Q^2) > 0.
\label{relatqdqbar}
\end{equation}
This means that the flavor asymmetry of the light antiquark distributions is the 
same for the corresponding helicity distributions, as noticed long time ago 
\cite{bbs-rev} (see also ref. \cite{sala17}).\\
Now let us come back to all these components 
$xu_+ (x, Q^2), ...x\bar u_- (x,Q^2)$ and more precisely to their $x$-behavior. 
It is clear that $xu_+(x,Q^2)$ is the largest one and they are all 
monotonic decreasing functions of $x$ at least for $x > 0.2$, outside the region 
dominated by the diffractive contribution.\\
Similarly $x\bar d_- (x\,Q^2)$ is the largest of the antiquark components.\\
Therefore if one considers the ratio $d(x,Q^2)/u(x,Q^2)$, its value is one at 
$x=0$, because the diffractive contribution dominates and, due to the monotonic 
decreasing property, it decreases for an increasing $x$.\\
This falling $x$-behavior has been verified experimentaly from the ratio of the DIS 
structure functions $F_{2}^{d} / F_{2}^p$ and the charge asymmetry of the 
$W^{\pm}$ production in $\bar p p$ collisions \cite{Kuhlmann:1999sf}.\\
Similarly if one considers the ratio  $\bar u(x,Q^2) /\bar d(x,Q^2)$, its value is 
one at $x=0$, because the diffractive contribution dominates and, due to the 
monotonic decreasing property, it also decreases for an increasing $x$.\\
By looking at the curves of Figure \ref{ratios}, one sees similar behaviors. 
In both cases in the vicinity of $x=0$ one has a sharp behavior due to the fact 
that the diffractive contribution dominates and in the high $x$ region
there is a flattening out above $x \simeq 0.6$. It is remarkable to observe that 
these ratios have almost no $Q^2$ dependence.

To conclude one predicts a monotonic increase of the ratio 
$\bar d (x,Q^2) /\bar u(x,Q^2)$. This was first observed in the low $x$ region by 
the E866/NuSea collaboration \cite{E866a}-\cite{E866b} and very recently there is a serious 
indication
from the preliminary results of the SeaQuest collaboration, that this trend persists
 beyond $x=0.2$ \cite{reimer}.
\section{ A toy model  for gluon distributions }
\label{toymodel}
In the BS15 version of the model \cite{Bourrely:2015kla},
the unpolarized gluon is parametrized 
as a Bose-Einstein function with a zero potential value and no diffractive 
part is included:
\begin{equation}
xG(x,Q^2_0) = \frac{A_Gx^{b_G}}{\exp(x/\bar{x})-1}~,
\label{eq3}
\end{equation}
where $A_G = 36.778$ is determined by the momentum sum rule.
The polarized gluon  distribution involves also a Bose-Einstein function
but requires an extra factorized function
whose origin is discussed in Refs. \cite{bourr16,bs2015}, so its expression
 is given by:
\begin{equation}
 x\Delta G(x,Q^2_0) = \frac {\tilde A_G x^{\tilde b_G}}{(1+ c_G
x^{d_G})}\!\cdot\!\frac{1}{\exp(x/\bar x) - 1 } \,,
\label{deltaG}
 \end{equation}
where $\tilde {A}_G = 26.887$. Contrary to the quarks situation these expressions
are not directly related and so have to be determined independently from 
specific experimental data.
Coming back to the model structure
this is not exactely true because their determination is influenced by 
 unpolarized and polarized quarks which are related, nevertheless, a more
direct relation will reinforce the model structure.

Inside a proton at high energy  beside the presence of 2 u + d quarks
there exists a collection of $q-\bar q$ pairs and gluons.
It is also know that a quark-antiquark pair can annihilate into 2 gluons.
It seems natural to suppose that a $q-\bar q$ pair
should behave like a composite boson and so could have a relation
with the gluon field. In this case one should
find that in a QCD process involving gluons, for instance in structure functions,
 one can replace the gluon by a $q-\bar q$ pair, leading to a new test
for the antiquarks since the quarks are well established.\\
For this purpose two new formulas are defined for the unpolarized and
polarized gluon which play the role of a toy model at the input scale.
In these formulas  $q$ and $\bar q$ contain only
the non diffractive part of Eqs. (\ref{eq1}, \ref{eq2}) and to comply with
 the previous definitions (\ref{eq3}, \ref{deltaG}), their expressions are now given by:
\begin{eqnarray}
xG(x,Q^2_0) &=& A_{q \bar q}(xu x\cdot x\bar u 
+ xd\cdot x\bar d + xs\cdot x\bar s)[x,Q^2_0]\,,
\label{gqqbar}\\
 x\Delta G(x,Q^2_0) &=&  A_{\delta q\bar q}
(x\Delta u\cdot x\Delta  \bar u + x\Delta  d \cdot x\Delta  \bar d 
+ x\Delta  s \cdot x\Delta \bar s)[x,Q^2_0]\,.
\label{dgqqbar}
\end{eqnarray}

Let us remark that the two formulas (\ref{gqqbar},\ref{dgqqbar}) although
they contain the product of 2 Fermi functions  both are 
evolved  as a  boson, so the 
result is not the evolution of the product of two Fermi distributions.\\
Also, in the expressions (\ref{eq3}, \ref{deltaG}) $G$ and $\Delta G$ are 
defined independently and are not related, while in the expressions
(\ref{gqqbar},\ref{dgqqbar}) indeed they are because for a given flavour
$q, \bar q, \Delta q,  \Delta \bar q$ are not independent. 
It has the consequence that the parton structure
can be described with a very few number of basic constituents.\\

A fit at NLO of unpolarized and polarized DIS experimental data
gives

\noindent in the case of BS15 parametrization \cite{Bourrely:2015kla}
\beq
\chi^2 = 2860 \quad\quad 2140 pts \quad\quad 1.34 \chi^2/pt
\label{solut1}
\eeq

now with Eqs. (\ref{gqqbar},\ref{dgqqbar}) of the toy model a fit of the same
set of data gives:
\beq
\chi^2 = 3013 \quad\quad 2140 pts \quad\quad 1.4 \chi^2/pt\,,
\label{solut2}
\eeq
the difference in $\chi^2$ is 5\%. 
Restricted to the polarized structure functions $g_1^p, g_1^d, g_1^n$
with 271pts, BS15 gives a $\chi^2 = 323$, the toy model a $\chi^2 = 301$.
Notice that in the original version the expression of  $\Delta G$ 
requires 4 parameters, in this version only one normalization constant 
$ A_{\delta q \bar q}$ is necessary, 
since $A_G,~ A_{q \bar q}$ are determined by the momentum sum rule,
this difference confirms the interest for the gluons given by
 Eqs. (\ref{gqqbar},\ref{dgqqbar})

In this new fit the potentials read
\begin{eqnarray}
&& X_{u}^{+}  =   0.4616 ,\quad  X_{u}^{-} =   0.3166,  \nonumber  \\     
&& X_{d}^{+} =   0.2530,\quad  X_{d}^{-} =   0.3062,  \nonumber  \\
&& X_{s}^{+} =   0.007896,\quad  X_{s}^{-} = 00982 ,  \nonumber  \\ 
&& b_{q}     = 0.491 ,\quad \bar b_{q} = 1.123    ,  \nonumber  \\
&& b_{s}     = 0.0044 ,\quad \bar b_{s} = 0.08    ,  \nonumber  \\
&&  \bar x =    0.0944\,. 
\label{newpot}
\end{eqnarray}

The new result for the potential values are close to the previous
ones (\ref{potval}) and still satisfy  the 
previous hierarchy (\ref{potherar}).
\beq
X_{u}^{+} >  X_{d}^{-} \sim  X_{u}^{-} > X_{d}^{+}\,,
\label{eq5}
\eeq
so the properties discussed in sec. \ref{propert} remain valid.

For the normalization constants one obtains
$A_{q \bar q} = 23.882$, $A_{\delta q \bar q} = 18.99$.

\begin{figure}[htp]   
\begin{center}
\includegraphics[width=8.0cm]{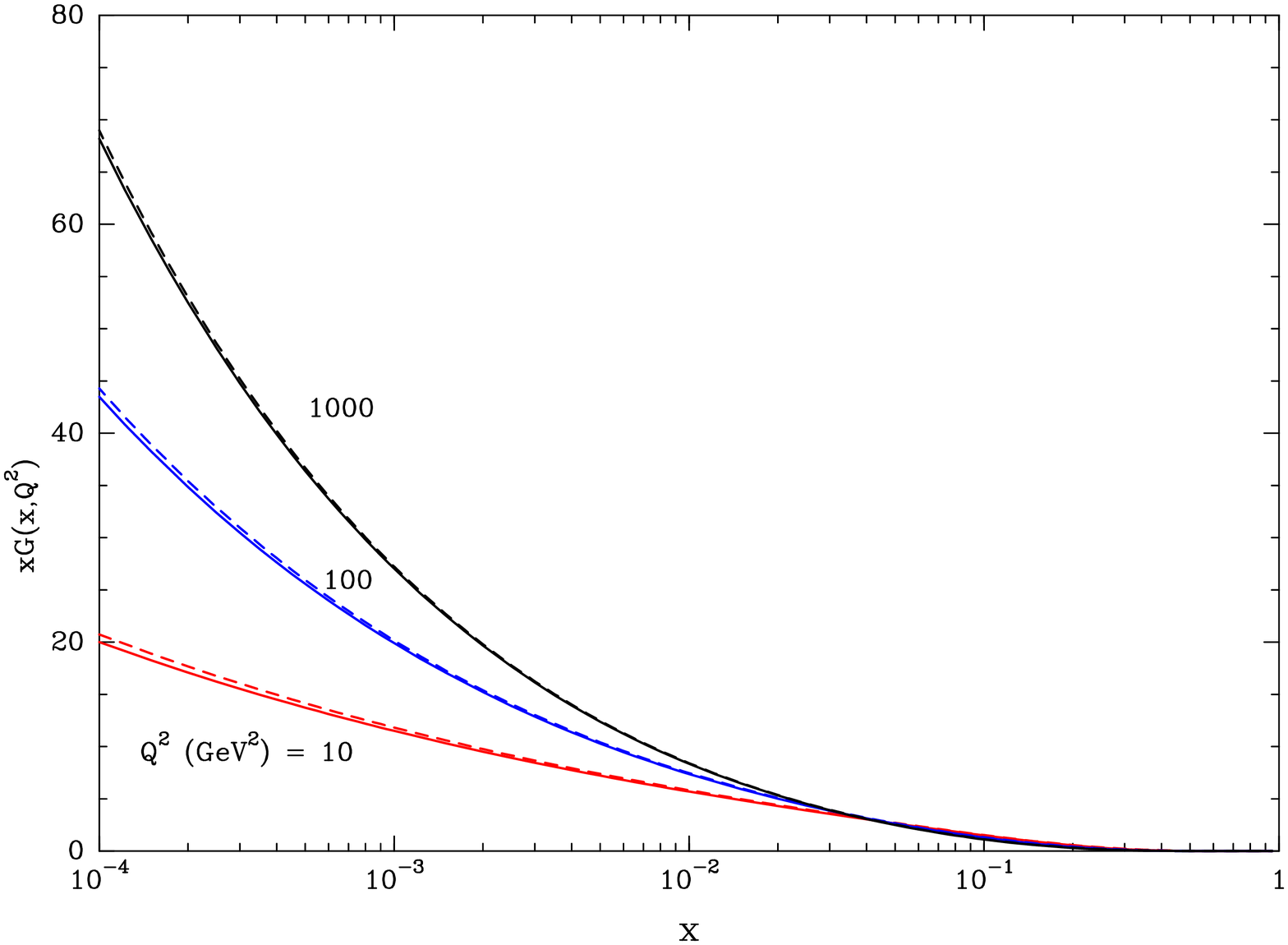}
\vspace*{+7.0ex}
\includegraphics[width=8.0cm]{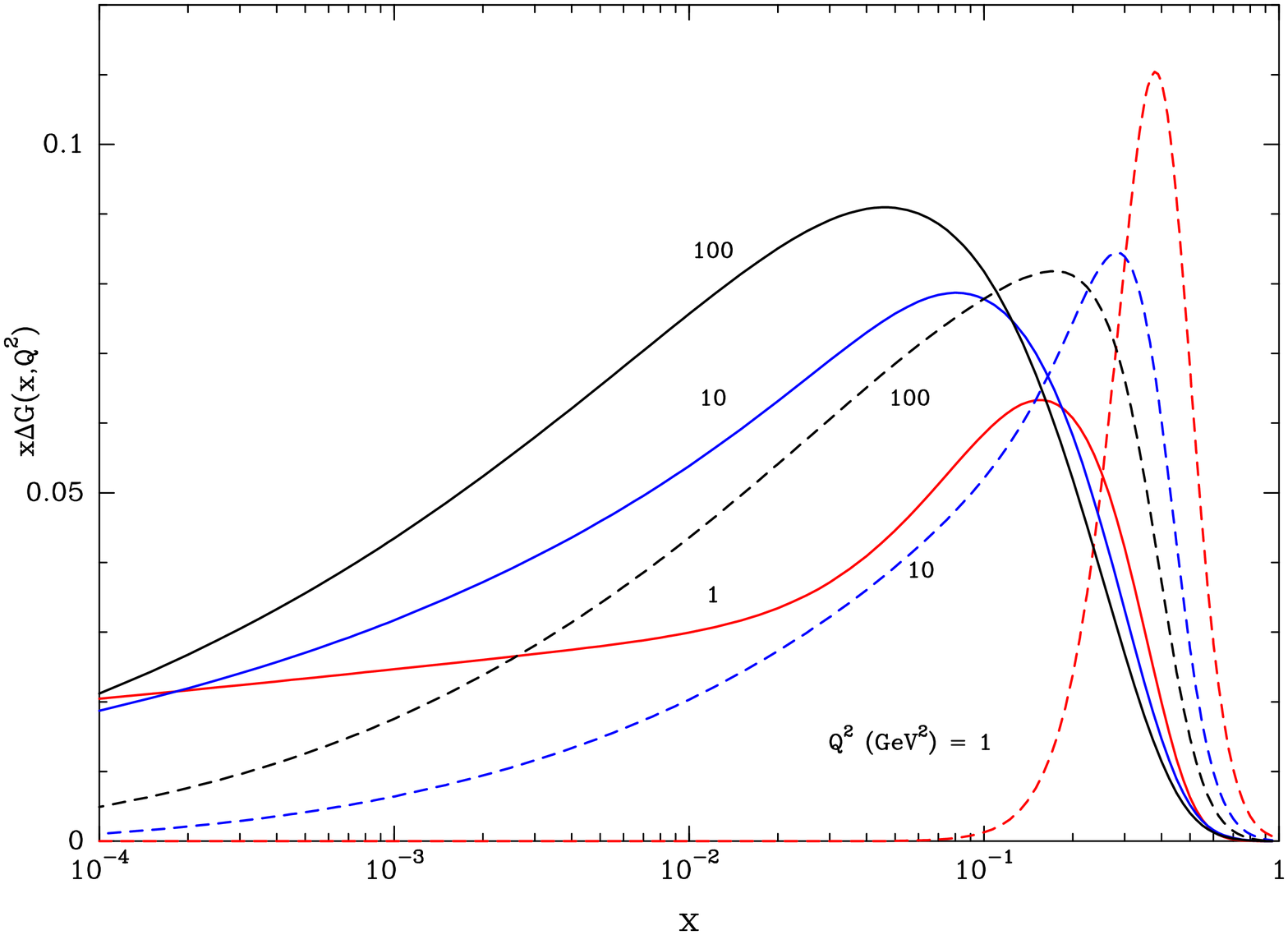}
\vspace*{-5.0ex}
\caption[*]{\baselineskip 1pt
Comparison of $xG(x,Q^2)$ computed from BS15 \cite{Bourrely:2015kla} 
(dashed) and the toy model (solid) (${\it left}$). 
Same comparison for $x\Delta G(x,Q^2)$ (${\it right}$) versus $x$ 
for  $Q^2 = 1, 10, 100 \mbox{GeV}^2$.}
\label{toybs15}
\end{center}
\end{figure}

In Figure \ref{toybs15} a plot is given for some results associated with the unpolarized
and polarized gluon in the case of BS15 parametrization 
(\ref{eq3}, \ref{deltaG}) (dashed curves)
and the toy model (\ref{gqqbar}, \ref{dgqqbar}) (solid curves).
The distributions behavior is very similar, the polarized case which is
more sensitive to the gluon structure 
 looks slightly different  but when combined with the polarized
quarks give an excellent description of the polarized structure
functions (see the $\chi^2$ discussed above). 
To  conclude this part devoted to the statistical model, 
the present formulas  used 
as a toy parametrization of unpolarized and polarized gluons  
give an  equivalent
description of the original model, and they represent also a new test
for the antiquarks PDFs since the quarks PDFs are well established.
In QCD calculations Mellin transforms are sometime involved,
the Mellin transform of a Fermi function for fermions and bosons
are mathematically related \cite{srivast11}, which is an encouraging sign for
our new definition of gluons.

One can ask the question if the previous formulation can be applied to
another model.
In the domain of polarized PDFs the DSSV model \cite{dssv1} is a reference,
so it becomes of interest to test this polarized version inside the toy model 
 taking DSSV as input in Eq. (\ref{dgqqbar}).
There is a difference between the statistical model and DSSV model
due to the fact that in the  statistical model unpolarized and polarized
PDFs are related which is not the case with DSSV.
Polarized quarks, antiquarks and gluon of flavor $i$ are defined in DSSV at the 
input scale $\mu_0$ by the expressions Eq. (28) of  Ref.  \cite{dssv1} namely
\begin{equation}
\label{eq:pdf-input}
x\Delta f_i(x,\mu^2_0) = N_i x^{\alpha_i} (1-x)^{\beta_i}
(1+\gamma_i \sqrt{x}+\eta_i x)\,.
\end{equation}

More serious constraint on the polarized gluon can be obtained from
the double-spin asymmetry in jet production $A_{LL}^{jet}$ with
the modified expression for the polarized DSSV gluon \cite{dssv2}
\begin{equation}
x\Delta g(x,Q_0^2)=N_g x^{\alpha_g}(1-x)^{\beta_g}
\left(1+\eta_g x^{\kappa_g}\right)\,.
\label{dginp}
\end{equation}
In order to test the toy model with the  polarized gluon (\ref{dgqqbar})
one adopts the strategy to fit  the same polarized data
previously used taking Eqs. (\ref{eq:pdf-input}) for the quarks and 
Eq. (\ref{dgqqbar}) for the polarized gluon. For simplicity
 the quarks the number of free parameter is restricted to $N_i, \eta_i$, 
while $\alpha_i, \beta_i, \gamma_i, $ are held fixed to their original values
(see Table II of  Ref.  \cite{dssv1}).

\begin{table}[ht!]
\begin{center}
\begin{tabular}{cccccc}
\hline
flavor $i$ &$N_i$ & $\alpha_i$ & $\beta_i$ &$\gamma_i$ &$\eta_i$\\
\hline
$u+\bar{u}$ & 0.403  & 0.692 & 3.34 & -2.18 & 21.38      \\
$d+\bar{d}$ & -0.023 & 0.164 & 3.89 & 22.40 & 83.80\\
$\bar{u}  $ & 4.83  & 0.692 & 10.0 & 0     & 24.97      \\
$\bar{d}  $ & -0.147 & 0.164 & 10.0 & 0     & 98.94      \\
$s = \bar{s}  $ & -0.019 & 0.164 & 10.0 & 0     & -23.03     \\
\hline
\end{tabular}
\caption { Parameters
describing NLO ($\overline{\mathrm{MS}}$)
$x\Delta f_i$ in Eq.~(\ref{eq:pdf-input})
at the input scale $\mu_0=1\,\mathrm{GeV}$, using the toy model}
\label{tab:para}
\end{center}
\end{table}

\begin{figure}[htp]   
\begin{center}
\includegraphics[width=9.0cm]{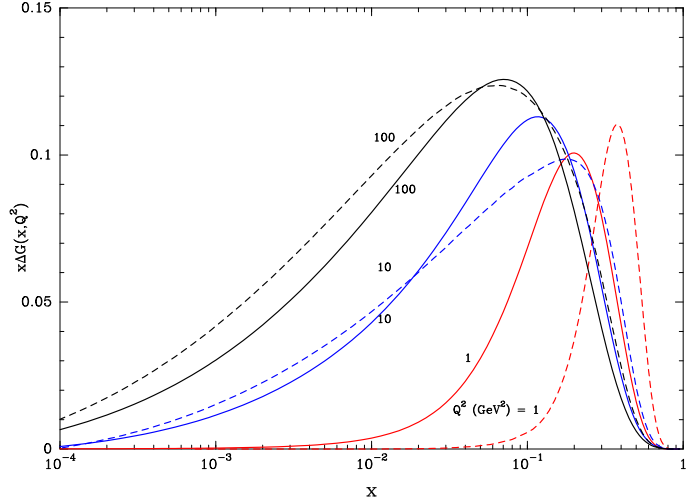}
\caption[*]{\baselineskip 1pt
Comparison of $x\Delta G(x,Q^2)$  versus $x$ 
for  $Q^2 = 1, 10, 100 \mbox{GeV}^2$ calculated with
the toy DSSV (solid) and the orignal one (dashed).}
\label{toydssv}
\end{center}
\end{figure}

For the polarized gluon one obtains a normalization coefficient 
 $A_{\delta q \bar q} = -0.078$. 
With a $\chi^2 = 235$ for 271pts  the quality of the polarized fit is 
similar to the previous statistical model. 
Here again  the five parameters
introduced in Eq. (\ref{dginp}) are reduced to one.
A plot of the polarized gluon for three $Q^2$ values
is shown in Fig. \ref{toydssv} for the original DSSV model (dashed curve)
and the toy model (solid curve).

\begin{figure}[htp]   
\begin{center}
\includegraphics[width=5.0cm]{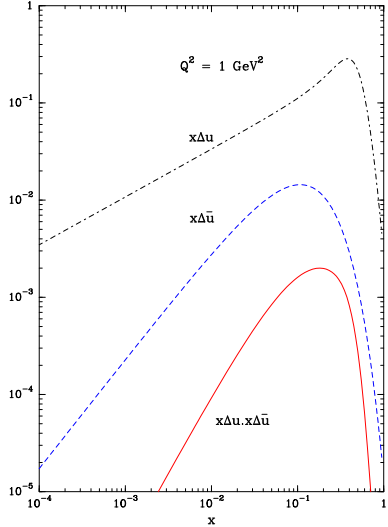}
\caption[*]{\baselineskip 1pt
  Quark $u$ contribution to the polarized $\Delta G$ following the toy
expression (\ref{dgqqbar}).}
\label{gltoy}
\end{center}
\end{figure}

Now our purpose is to show that 
the polarized gluon  discussed above offers a good exploratory domain
for the parton structure. Beginning with the statistical model, it was natural
to associate to the gluon a Bose-Einstein expression such that
\begin{equation}
 x\Delta G(x,Q^2_0) = \tilde A_G x^{\tilde b_G}
\!\cdot\!\frac{1}{\exp(x/\bar x) - 1 } \,,
\label{deltaG1}
 \end{equation}
this original expression was unable to describe
the double-spin asymmetry of the one-jet inclusive production $A_{LL}^{jet}$ in the near
forward rapidity region as a function of $p_T$ within the domain 
 5 $ \leq p_T \leq$ 30GeV measured by the STAR Collaboration at BNL-RHIC
 \cite{abelev06}.
To obtain a good description the polarized gluon was modified according to Eq. (\ref{deltaG}).\\
It turns out that the extra multiplicative function $\frac{1}{(1+ c_Gx^{d_G})}$
has the behavior of a logistic function or activation function used in neural
network \cite{bourr16}
\begin{equation}
S(x) = \frac{1}{1 +e^{-e_G x + h_G}}\,,
\label{sigmv1}
\end{equation}
so one can write the polarized gluon as:
\begin{equation}
 x\Delta G(x,Q^2_0) =  S(x) \frac{\tilde {A'}_G x^{b_G}}{\exp(x/\bar{x})-1}\,.
\label{polglues1}
 \end{equation}
The physical interpretation of this new formula means that
the incoming momentum is collected now by means  of a Bose-Einstein distribution and 
then filtered by an activation function to produce the gluon probability 
distribution. \\
The toy model defined above proceeds along the same line,
a polarized gluon is built in terms of a composite
made of known physical functions namely the PDFs associated with their
probability. 
In Fig. \ref{gltoy} the  example of $u$, $\bar u$ quarks
where their probabilities product generates a component of the
gluon polarized PDF.
The resulting effect of the toy model
 is perfectly compatible with experimental data for
both  the statistical and DSSV models.
\\
To summarize the discussion on the different expressions so far 
defined in (\ref{dgqqbar}),(\ref{deltaG1})-(\ref{polglues1})  our
objective was to replace an arbitrary function by a physical quantity
perfectly justified in the context of the model.\\
One knows that $\Delta G$ gives an important contribution to the proton spin sum rule.
A study of this effect is presented in Fig. 3 of Ref. \cite{bs2015} using the gluon
defined by Eq. (\ref{deltaG}). One sees that just above $Q^2 = 100 \mbox{GeV}^2$ 
the value of the spin sum rule 1/2 is saturated, the same calculation performed with the toy gluon Eq. (\ref{dgqqbar}) gives a saturation for $Q^2$ around 1000GeV$^2$, 
which corresponds to a significant improvement.

\begin{figure}[htp]   
\vspace*{-5.0ex}
\begin{center}
\includegraphics[width=5.9cm]{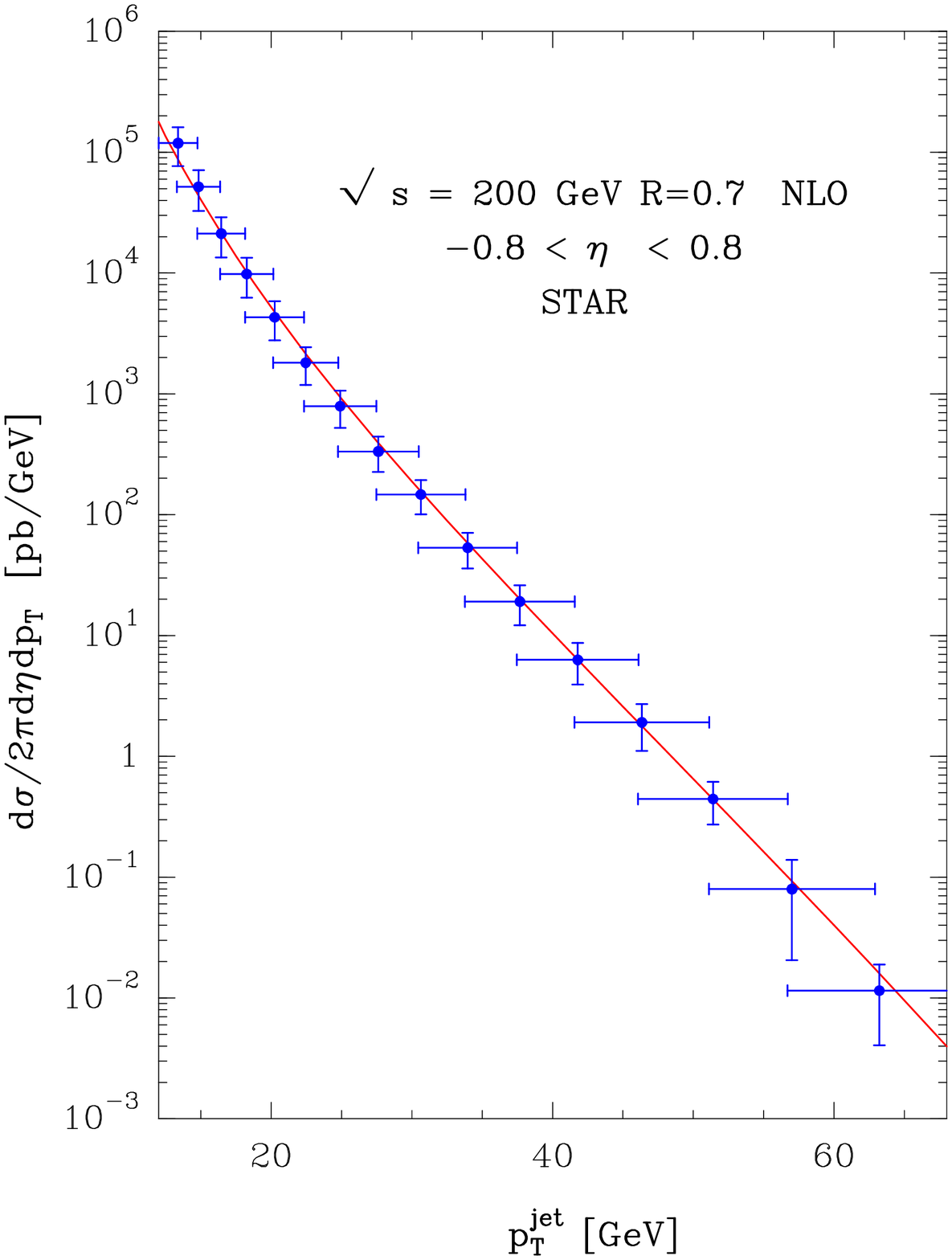}
\includegraphics[width=6.0cm]{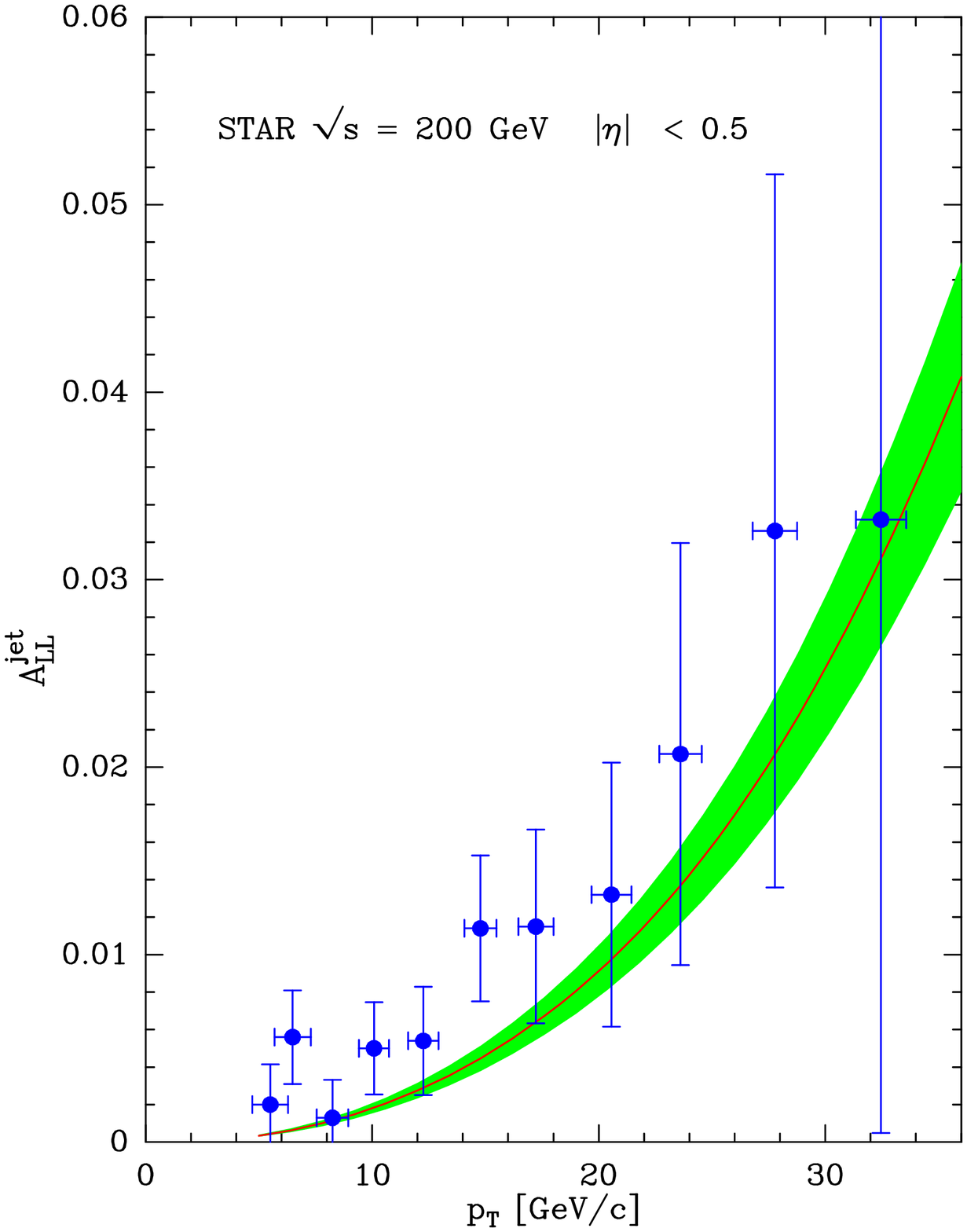}
\vspace*{1.0ex}
\caption[*]{\baselineskip 1pt
 [{\it left}]
Inclusive differential cross section for $p + p \rightarrow jet + X$
at $\sqrt{s}$ =  200 GeV versus jet $p_T$ calculated with the 
unpolarized toy gluon Eq. (\ref{gqqbar}).  [{\it right}]
The longitudinal double-spin asymmetry $A_{LL}$ in 
$\overrightarrow{p} + \overrightarrow{p} \rightarrow jet + X$
at $\sqrt{s}$ =  200 GeV versus jet $p_T$ calculated with the 
polarized toy gluon Eq. (\ref{dgqqbar}). Data STAR experiment \cite{Adamczyk:2014xyw}.}
\label{jettoy}
\end{center}
\end{figure}

Finally, one would like to present a new test of the toy gluon  distribution in a
pure hadronic reaction and compute the double-helicity asymmetry $A^{jet}_{LL}$
discussed above. It is important to remark that the asymmetry calculation requires both the knowledge of the unpolarized and polarized gluon distributions 
(\ref{gqqbar})-(\ref{dgqqbar}).
In Fig. \ref{jettoy} our  prediction is
compared with these high-statistics data points and the agreement is very reasonable.

\clearpage
\newpage
\section{ Conclusion}
Our purpose was to show that a statistical model offers a unique framework to build 
quarks structure  whose properties are clearly defined by parameters related to physical
quantities in the PDFs expressions. The thermodynamical potential 
which are the master parameters generate definite
properties of the quarks PDFs  confirmed by experimental structure functions.

This prediction results from the following characteristic features of the 
statistical approach:\\
- the PDF helicity components defined by Fermi-Dirac expressions are
the building blocks of the unpolarized and polarized PDFs.\\
-  the thermodynamical potentials satisfy
a hierarchy relation given by  Eq. (\ref{potherar}) which imposes
specific properties on the distribution functions.\\
-  the expressions between quark and antiquarks 
obtained  allow to relate the behavior of the ratios  
$xd(x,Q^2)/xu(x,Q^2)$ and $x\bar u(x,Q^2) /x\bar d(x,Q^2)$.\\
-  a toy model has been defined for the gluon in terms of unpolarized and
polarized quarks distributions  which produces equivalent results to the original
gluon parametrizations but with only one free normalization parameter. 
In addition this toy model gives for the gluon  made with basic
fermion helicity components a relation
between unpolarized and polarized gluons distributions which was not the
case in the original version of the model.

It is clear that our model is able to explain a large set of 
unpolarized and polarized experimental Deep Inelastic Scattering data.
Of course the predictions which can be made in view of future experiments depend
on the present values of the parameters so it is a challenge
for the model to be confimed by new experiments.

To conclude our statistcal approach not only provides numerical PDFs  values
compatible with experimental data but also gives a coherent model of the quarks
structure at the fundamental level of helicity distributions.


\end{document}